\documentclass[a4paper,twocolumn,superscriptaddress,pre, aps, floatfix ]{revtex4-1}
\usepackage{amssymb}
\usepackage{amsmath}
\usepackage{marvosym}
\usepackage{graphicx}
\usepackage{color}
\usepackage{url}
\usepackage{lipsum}
\usepackage{soul}
\usepackage{multirow}
\usepackage[colorlinks=true,allcolors=blue]{hyperref}

\renewcommand{\epsilon}{\varepsilon}
\newcommand{\R}{\mathbb{R}}

\begin{document}

\title{Small worlds and clustering in spatial networks}

\author{Mari\'an Bogu\~na}
\email{marian.boguna@ub.edu}
\affiliation{Departament de F\'isica de la Mat\`eria Condensada, Universitat de Barcelona, Mart\'i i Franqu\`es 1, E-08028 Barcelona, Spain}
\affiliation{Universitat de Barcelona Institute of Complex Systems (UBICS), Universitat de Barcelona, Barcelona, Spain}

\author{Dmitri Krioukov}
\email{dima@northeastern.edu}
\affiliation{Network Science Institute, Northeastern University, 177 Huntington avenue, Boston, MA, 022115} 
\affiliation{Department of Physics, Department of Mathematics, Department of Electrical~\&~Computer Engineering, Northeastern University, 110 Forsyth Street, 111 Dana Research Center, Boston, MA 02115, USA}

\author{Pedro Almagro}
\affiliation{Departament de F\'isica de la Mat\`eria Condensada, Universitat de Barcelona, Mart\'i i Franqu\`es 1, E-08028 Barcelona, Spain}
\affiliation{Universitat de Barcelona Institute of Complex Systems (UBICS), Universitat de Barcelona, Barcelona, Spain}

\author{M. \'Angeles \surname{Serrano}}
\email{marian.serrano@ub.edu}
\affiliation{Departament de F\'isica de la Mat\`eria Condensada, Universitat de Barcelona, Mart\'i i Franqu\`es 1, E-08028 Barcelona, Spain}
\affiliation{Universitat de Barcelona Institute of Complex Systems (UBICS), Universitat de Barcelona, Barcelona, Spain}
\affiliation{Instituci\'o Catalana de Recerca i Estudis Avan\c{c}ats (ICREA), Passeig Llu\'is Companys 23, E-08010 Barcelona, Spain}

\begin{abstract}
Networks with underlying metric spaces attract increasing research attention in network science, statistical physics, applied mathematics, computer science, sociology, and other fields. This attention is further amplified by the current surge of activity in graph embedding. In the vast realm of spatial network models, only a few reproduce even the most basic properties of real-world networks. Here, we focus on three such properties---sparsity, small worldness, and clustering---and identify the general subclass of spatial homogeneous and heterogeneous network models that are sparse small worlds and that have nonzero clustering in the thermodynamic limit. We rely on the maximum entropy approach where network links correspond to noninteracting fermions whose energy dependence on spatial distances determines network small worldness and clustering.
\end{abstract}

\maketitle

In spatial networks, nodes are positioned in a geometric space, and the distances between them in the space affect their linking probability in the network~\cite{barthelemy2011spatial}. In real-world systems, such spaces can be explicit/physical, as in geographically embedded networks~\cite{latora2002boston,guimera2005worldwide} or in the Ising model with long-range interactions~\cite{gitterman2000small,viana-lopes2004exact,teles2012nonequilibrium}. Yet these spaces can be also hidden/latent. Latent similarity spaces have been employed for nearly a century to model homophily in social networks, for instance~\cite{sorokin1927social,majone1972social}: the closer the two people are in a virtual similarity space, the more similar they are, the more likely they know each other~\cite{mcpherson2001birds}. Another field where the space can be virtual are graph embeddings in computer science and machine learning, with applications including network compression, visualization, and node labeling~\cite{grover2016node2vec,goyal2018graph}. 

In models of spatial networks, the space is usually explicit. Perhaps the simplest spatial network model is that of random geometric graphs that have been extensively studied in mathematics and physics since the early 60ies~\cite{gilbert1961random,penrose2003random,dall2002random,coon2017entropy}. In these graphs, nodes are positioned in a space randomly using a point process, usually a Poisson point process, and two nodes are linked in the graph if the distance between them in the space is less than a fixed threshold. If the intensity of the point process does not depend on the graph size~$n$, then the resulting graphs are sparse and have nonzero clustering in the thermodynamic $n\to\infty$ limit, thus sharing these two properties with many real-world complex networks~\cite{barabasi2016network,newman2018networks}. Yet many of these networks are also heterogeneous small worlds, while random geometric graphs are homogeneous large worlds.

This mismatch was resolved in~\cite{serrano2008similarity,krioukov2010hyperbolic} where a class of models of spatial networks that are sparse heterogeneous small worlds with nonzero clustering was introduced. Networks in these models have some additional properties commonly observed in real-world networks, such as self-similarity~\cite{serrano2008similarity,garcia-perez2018multiscale} and community structure~\cite{zuev2015emergence,muscoloni2018nonuniform,garcia-perez2018soft}. Yet the following question remains: what are the general requirements to spatial network models so that networks in these models possess the properties of real-world networks?

Here, we first focus on just three properties: (1)~sparsity, (2)~small worldness, and (3)~nonzero clustering. Simplifying the results a bit, we show that spatial networks in~$\R^d$ have all these three properties at once only if the probability~$p_{ij}$ of connection between nodes~$i$ and~$j$ scales with the distance~$x_{ij}$ between them in~$\R^d$ as $p_{ij}\sim x_{ij}^{-\beta}$ with $\beta\in(d,2d)$. We then add (4)~heterogeneity to the list of the requirements, and show that~$\beta$ must be within the same range~$(d,2d)$ if the variance of the degree distribution is finite. If it is infinite, however, e.g.\ if it is a power law with exponent $\gamma\in(2,3)$, then the networks are always ultrasmall worlds, and any $\beta>d$ satisfies all the four requirements. Finally, we show that if we also want to suppress nonstructural degree correlations, then the unique shape of the connection probability in the heterogeneous case is as in~\cite{serrano2008similarity,krioukov2010hyperbolic}: $p_{ij} \sim (\kappa_i\kappa_j)^{\beta/d}x_{ij}^{-\beta}$, where $\kappa_i,\kappa_j$ are the expected degrees of nodes~$i,j$.

To obtain these results, we take a statistical physics stance in which we interpret spatial network models as probabilistic mixtures of grand canonical ensembles that maximize ensemble entropy under certain constraints, and are thus statistically unbiased. We call these mixtures \emph{hypergrandcanonical ensembles}, as some of their parameters are random, and random parameters are known as \emph{hyperparameters} in statistics.

{\bf Settings and notations.} We consider a very general class of spatial network models. The space is any compact homogeneous and isotropic Riemannian manifold of dimension~$d$ and volume~$n$, and with no boundaries. We require the curvature of the manifold to go to zero at $n\to\infty$. That is, the space is locally the Euclidean space, and it is exactly the Euclidean space~$\R^d$ in the thermodynamic limit. Examples are the $d$-sphere or $d$-torus of size growing with~$n$ such that its volume is~$n$. Any growing compact $d$-dimensional hyperbolic manifold with no boundaries is also fine. On such a manifold we sprinkle $n$ points uniformly at random according to the manifold metric. These points are thus the binomial point process of rate~$1$ on the manifold, and they form the node set of a random graph. Conditioned on node coordinates on the manifold, nodes $i$ and $j$ are connected independently with probabilities $p_{ij}=p(x_{ij})$, where $x_{ij}$ is the distance between $i$ and $j$ on the manifold. By $a_{ij}$ we denote the adjacency matrix of these random graphs: conditioned on node coordinates, $a_{ij}$s are independent Bernoulli random variables with success rates~$p_{ij}$. These graphs are known as soft random geometric graphs~\cite{penrose2016connectivity,dettmann2016random}.

We interpret these random graph ensembles as probabilistic mixtures of grand canonical ensembles that maximize ensemble entropy under the constraints that the average number of particles and average energy are fixed to given values. Particles are edges~$a_{ij}$ here, and their energies~$\epsilon_{ij}$ depend on distances~$x_{ij}$: $\epsilon_{ij}=f(x_{ij})$. The connection probability function~$p(x)$ then takes the familiar Fermi-Dirac form, see Sec.~\ref{app1},
\begin{equation}\label{eq:fd}
  p(x_{ij}) = \frac{1}{e^{\beta(f(x_{ij})-\mu)}+1}.
\end{equation}
The Lagrange multipliers corresponding to the number-of-particles and energy constraints are the chemical potential~$\mu$ and inverse temperature~$\beta\geq0$, as usual. We assume that neither $f(x)$ nor $\beta$ depend on~$n$, but $\mu$, and consequently the absolute activity $\lambda = e^{\beta\mu}$, can depend on~$n$ as they usually do in statistical physics. Since energies~$\epsilon_{ij}$ are not fixed as in grand canonical ensembles but are random instead, we call this ensemble a \emph{hypergrandcanonical ensemble}.

We require our networks to be always \emph{sparse}, meaning that the expected average degree in them is fixed to a finite positive constant~$\langle k \rangle$ for any network size~$n$. We call a network model a \emph{small world} if the average hop distance of shortest paths in the model networks grows slower than any polynomial of $n$. In particular, average distances growing as any polynomial of $\ln{n}$ in a model would render the model a small world. The model is also an \emph{ultrasmall world} if the average distance grows slower than any polynomial of $\ln{n}$. If a model is not a small world, we call it a \emph{large world}. By \emph{clustering} we mean the average local clustering coefficient. Symbol~`$\sim$' in $a_n \sim b_n$ or $a(x) \sim b(x)$ means that $a_n/b_n$ or $a(x)/b(x)$ converge to a finite positive constant at $n\to\infty$ or $x\to\infty$, respectively.

{\bf Homogeneous spatial networks.} In any network model satisfying the settings above, the degree distribution is homogeneous: in the thermodynamic limit $n\to\infty$ it converges to the Poisson distribution with the mean equal to the average degree in the network~\cite{boguna2003class,bollobas2007phase}. By \emph{network homogeneity} we mean here not only degree homogeneity, but also all the consequences of the manifest invariance of these ensembles with respect to the group of isometries of the manifold. In particular, the expected values of any graph property of any two nodes in these random graphs are the same. For instance, not only the expected degree, but also the expected clustering of any two nodes is the same and equal to the average clustering in the network. The main question is under what conditions these networks are small worlds and have nonzero clustering in the thermodynamic limit. 

\begin{table}
\begin{tabular}{|l|l|l|}
  \hline
  Parameter regime & Small world & Clustering \\
  \hline\hline
  $\beta\to0$ (ER)  & \multirow{2}{*}{Yes} & \multirow{2}{*}{No} \\
  $\beta < d/l_{\sup}$  & & \\
  \hline
  $d/l_{\inf} < \beta < 2d/l_{\sup}$ & Yes & Yes \\
  \hline
  $\beta > 2d/l_{\inf}$ & \multirow{2}{*}{No} & \multirow{2}{*}{Yes} \\
  $\beta\to\infty$ (RGG) & & \\
  \hline
\end{tabular}
\caption{
{\bf The result summary for homogeneous networks.} \emph{Small world}: yes/no: the networks are small/large worlds. \emph{Clustering}: yes/no: the networks have nonzero/zero clustering in the thermodynamic ($n\to\infty$) limit. \emph{ER}: Erd\H{o}s-R\'enyi random graphs~\cite{erdos1959random}. \emph{RGG}: sharp random geometric graphs in~$\R^d$~\cite{gilbert1961random}. \emph{Parameters}: $\beta$ the inverse temperature, $d$ the space dimension, $l_{\inf} = \liminf_{x\to\infty}f(x)/\ln{x}$, $l_{\sup} = \limsup_{x\to\infty}f(x)/\ln{x}$, where $f(x)$ is the energy function: $\epsilon_{ij} = f\left(x_{ij}\right)$, where $x_{ij}$ is the distance between nodes $i$ and $j$ in the space. Note that $l_{\sup}=0$ corresponds to $f(x)$ growing slower than logarithmically, in which case the networks are in the first regime for any value of $\beta<\infty$. Note that $l_{\inf}=\infty$ corresponds to $f(x)$ growing faster than logarithmically, in which case the networks are in the last regime for any value of $\beta>0$. Note that $f(x) = c\,\ln{x} + o\left(\ln{x}\right)$ corresponds to $l_{\inf}=l_{\sup}=c$. The cases with $\beta\in[d/l_{\sup},d/l_{\inf}]$, $\beta\in[2d/l_{\sup},2d/l_{\inf}]$, and $l_{\sup}\geq2l_{\inf}$ require further details about the specific shape of $f(x)$ to classify the network into one of the three shown classes.
\label{tab:homo}
}
\end{table}

The results are summarized in Table~\ref{tab:homo}. Intuitively, they are easy to comprehend. If $f(x)$ grows too fast with~$x$, so that $p(x)$ decays too fast, then the network does not have sufficiently long links that are needed for small worldness. The network is thus necessarily a large world. On the other hand, if $f(x)$ grows too slow with~$x$, so slow that with an $n$-independent~$p(x)$ the average degree diverges, then to have an $n$-independent average degree the absolute activity~$\lambda$ must depend on~$n$ and go to zero at $n\to\infty$, meaning that $p_{ij}$s go to zero as well. But since clustering scales with $n$ the same way as $p_{ij}$s do (recall that average clustering is the probability that two random neighbors of a random node are connected), it is zero at $n\to\infty$. Luckily, there exists a ``sweet spot'' at which the rate of growth of $f(x)$ is not too fast and not too slow, so that the networks are small worlds and have nonzero clustering at the same time, the second regime in Table~\ref{tab:homo}.

To show that this sweet spot (or range indeed) is as shown in Table~\ref{tab:homo}, we first observe that the average degree in our graphs is
\begin{align}\label{eq:kbar-homo}
\langle k \rangle &= \left\langle k_i \right\rangle = \left\langle\sum_{j}a_{ij}\right\rangle
\sim \int^{n^{1/d}}x^{d-1}p(x)\,dx \nonumber\\
&\sim \lambda\int^{n^{1/d}}x^{d-1}e^{-\beta f(x)}dx=\lambda I_n.
\end{align}
This is because the number of nodes at distances~$[x,x+dx]$ from a given node~$i$ in~$\R^d$ is proportional to $x^{d-1}dx$, node~$i$ is connected to each of those nodes with probability~$p(x)$, and we integrate up to the space diameter, which is~$\sim n^{1/d}$. The lower integration limit is any positive constant.

If the integral~$I_n$ diverges with~$n$, then $\lambda$ must depend on~$n$ and go to zero at $n\to\infty$ to yield an $n$-independent~$\langle k \rangle$ above. But if $\lambda$ tends to zero, then $p_{ij}\sim \lambda e^{-\beta\epsilon_{ij}}$ tends to zero as well, and so does clustering. The integral $I_n$ diverges if the monotonic function $f(x)$ does not grow sufficiently fast. In particular, $I_n$ diverges if $l_{\sup}=\limsup_{x\to\infty}f(x)/\ln{x}<d/\beta$, the first regime in Table~\ref{tab:homo}. On the other hand, if $I_n$ converges---in particular, it does so if $l_{\inf}=\liminf_{x\to\infty}f(x)/\ln{x}>d/\beta$, the second row in Table~\ref{tab:homo}---then $\lambda$ is strictly positive, and so are $p_{ij}$s and clustering.

Turning to small worldness now, the network is a small world only if it contains links connecting nodes located at distances~$x_{ij}$ of the order of the space diameter~$\sim n^{1/d}$, as well as at all other smaller distances. Let $l(x)$ be the distribution of link lengths~$x$, defined as distances between linked nodes in the space. Observe that $l(x)\sim x^{d-1}p(x) \sim x^{d-1}e^{-\beta f(x)}$. If $f(x)=c\,\ln{x}$, then $l(x)\sim x^{-\delta}$ with exponent $\delta=c\beta-d+1$. Since networks are sparse, there are $\sim n$ links. The expected maximum value among $n$ samples from a power law with exponent~$\delta$ is $\sim n^\xi$ with $\xi=1/(\delta-1)$~\cite{boguna2004cut}. The network is a small world only if this expected maximum link length is larger than the space diameter $\sim n^{1/d}$, which implies $\xi>1/d$ or $\beta<2d/c$, cf.~the second row in Table~\ref{tab:homo} with $l_{\sup}=c$. If $f(x)$ grows faster than logarithmically, $l_{\inf}=\infty$, then $l(x)$ decays faster than a power law, $\xi=0$, and there are no long links at all, so that our networks are necessarily large worlds, the last regime in Table~\ref{tab:homo}.

\begin{figure}[!t]
	\centering
	\includegraphics[width=1\linewidth]{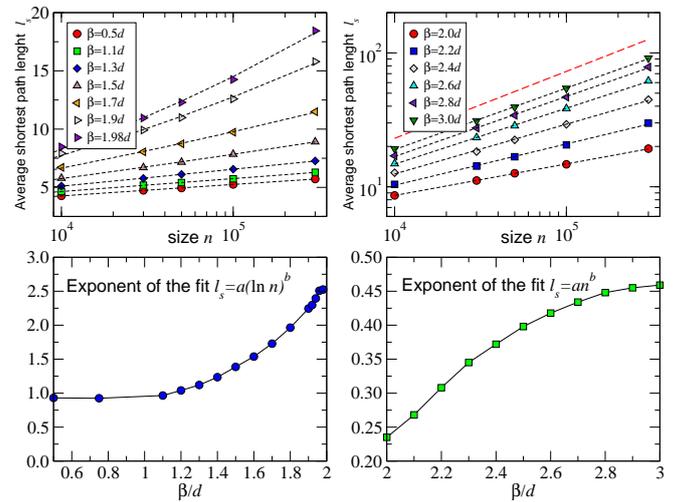}
	\caption{{\bf The average shortest path length~$l_s$ in the homogeneous spatial networks as a function of the network size~$n$.}
This function is measured for different values of inverse temperature $\beta$ in the connection probability~\eqref{eq:fd} with $f(x)=\ln x$ used to generate random networks on the $d=2$-dimensional sphere of area~$n$. The average degree in all these networks is fixed to $\langle k \rangle=10$ by the appropriate choice of the chemical potential~$\mu$, and the results are averaged over 10 random network realizations for each data point. The functions $l_s(n)$ are then fit with $a\ln^bn$ for $\beta<2d$ (left panels) and with $an^b$ for $\beta\geq2d$ (right panels), and the exponents $b$ of these fits as functions of $\beta/d$ are shown in the bottom panels. The dashed red line is $\sim n^{1/2}$, the distance scaling in the two-dimensional sharp RGGs corresponding to $\beta \rightarrow \infty$.
}
	\label{fig:1}
\end{figure}

This logic is about the necessary conditions for small worldness, but they have been proven to be also sufficient~\cite{benjamini2004geometry,biskup2019sharp,berger2004lower}, and we confirm all the results above in simulations in Fig.~\ref{fig:1} (small worldness) and in Fig.~\ref{fig:S2} (clustering) in the Sec.~\ref{app5}. Figure~\ref{fig:1} shows that the average shortest path length~$l_s$ scales with the network size~$n$ as $l_s\sim\ln^bn$ if $\beta<2d$, and as $l_s\sim n^b$ if $\beta\geq2d$. In the small world regime $\beta<2d$, the exponent~$b$ in $l_s\sim\ln^bn$ is close to~$1$ for any $\beta<d$, while for $\beta\in(d,2d)$ it is a growing function of $\beta$ that appears not to diverge but to approach some finite maximum value as $\beta$ approaches~$2d$. In the large-world regime $\beta\geq2d$, exponent~$b$ in $l_s\sim n^b$ is also growing function of $\beta$ ranging in values from some minimum value at $\beta=2d$ that does not appear to be zero, to its theoretical maximum $b=1/d$ at zero temperature $\beta\to\infty$ corresponding to sharp RGGs. The nature of the small-to-large world phase transition at $\beta=2d$ appears to be an interesting open question~\cite{biskup2019sharp}. The simulations can hardly reach network sizes that are sufficiently large to provide any hints regarding whether this transition is continuous or discontinuous, yet the results in Fig.~\ref{fig:1} suggest the latter since the continuous transition would yield small-world $b\to\infty$ at $\beta\to2d^-$ and large-world $b\to0$ at $\beta\to2d^+$.

{\bf Heterogeneous spatial networks.} Instead of the chemical potential~$\mu$, the Lagrange multiplier that fixes the expected average degree in the homogeneous ensemble, in the heterogeneous ensemble we have $n$ Lagrange multipliers~$\alpha_i$ that fix the expected degree~$\langle k_i \rangle = \left\langle\sum_{j}a_{ij}\right\rangle$ of each individual node to a desired value~$\kappa_i$. The relations between~$\kappa_i$ and~$\alpha_i$ are documented in Sec.~\ref{app3}. Here, we assume that the parameters~$\kappa_i$ are hyperparameters, meaning they are random and sampled from a fixed distribution~$\rho(\kappa)$, in which case we have the same hypergrandcanonical ensemble as in the homogeneous case above, except that the connection probability changes from~\eqref{eq:fd} to
\begin{equation}\label{eq:fda}
p(x_{ij},\alpha_i,\alpha_j)=\frac{1}{e^{\beta f(x_{ij})+\alpha_i+\alpha_j}+1}.
\end{equation}
The degree distribution in this ensemble converges to the mixed Poisson distribution $P(k)=(1/k!)\int_\kappa \kappa^ke^{-\kappa}\rho(\kappa)\,d\kappa$ whose shape ``follows'' the shape of $\rho(\kappa)$~\cite{boguna2003class,bollobas2007phase}. This type of heterogeneous spatial network models were first introduced in~\cite{serrano2008similarity}, and many other similar classes of models have been defined and studied since then~\cite{bonato2012geometric,deijfen2013scale,bringmann2019geometric}.

The qualitative behavior of clustering---zero versus nonzero in the thermodynamic limit---is exactly the same in these heterogeneous models as in the homogeneous one. Indeed, the expression~\eqref{eq:kbar-homo} for the average degree $\langle k \rangle$ changes to
\begin{align}\label{eq:kbar-hetero}
\langle k \rangle &\sim \int^{n^{1/d}}x^{d-1}\,dx \iint_{\alpha,\alpha'} p(x,\alpha,\alpha')\,\rho(\alpha)d\alpha\,\rho(\alpha')d\alpha' \nonumber\\
&\sim\left[\int_\alpha e^{-\alpha}\rho(\alpha)d\alpha\right]^2\int^{n^{1/d}}x^{d-1}e^{-\beta f(x)}dx=\hat{\lambda} I_n,
\end{align}
where $\hat{\lambda}=\langle e^{-\alpha} \rangle^2$, and $\rho(\alpha)$ is the distribution of Lagrange multipliers determined by the distribution of expected degrees~$\rho(\kappa)$. Following exactly the same reasoning as in the homogeneous case, albeit applied to~$\hat{\lambda}I_n$ instead of~$\lambda I_n$, we thus conclude that clustering is zero or nonzero at $n\to\infty$ depending on whether $I_n$ diverges or converges. For $f(x)=\ln x$ for example, this means that the situation is exactly the same as in the homogeneous case: the clustering is zero if $\beta<d$ and nonzero if $\beta>d$.

Turning to small worldness, we assume henceforth that $f(x) = \ln{x}$. We do so not only to simplify the discussion, but also because we prove in Sec.~\ref{app2} that $f(x)=\ln{x}$ is unique in the sense that this is the only possible form of $f(x)$ that does not induce any degree correlations other than the structural ones~\cite{boguna2004cut}. We also assume that the distribution~$\rho(\kappa)$ of expected degrees~$\kappa$ is the Pareto distribution
\begin{equation}\label{eq:pareto}
  \rho(\kappa)=(\gamma-1)\kappa_0^{\gamma-1}\kappa^{-\gamma}\text{, where }\kappa\geq\kappa_0>0\text{ and }\gamma>2.
\end{equation}
We note that the networks defined by~(\ref{eq:fda},\ref{eq:pareto}) with $f(x)=\ln{x}$ were introduced in~\cite{serrano2008similarity} and are equivalent to random hyperbolic graphs~\cite{krioukov2010hyperbolic}.

\begin{table}[t]
\begin{tabular}{|l|l|l|}
  \hline
  Parameter regime & $2<\gamma<3$ & $\gamma>3$, $\gamma=\infty$ \\
  \hline\hline
  $\beta\to0$ (HSCM)  & \multirow{2}{*}{USW, ZC} & \multirow{2}{*}{SW, ZC} \\
  $\beta < d$  & & \\
  \hline
  $d < \beta < 2d$ & USW, PC & SW, PC \\
  \hline
  $\beta > 2d$ & \multirow{2}{*}{USW, PC} & \multirow{2}{*}{LW, PC} \\
  $\beta\to\infty$ (RHG) & & \\
  \hline
\end{tabular}
\caption{
{\bf The result summary for the heterogeneous networks with $f(x)=\ln{x}$ and Pareto~$\rho(\kappa)$ as in~\cite{serrano2008similarity}.} The abbreviations are: \emph{HSCM}: the hypersoft configuration model~\cite{hoorn2018sparse}; \emph{RHG}: sharp random hyperbolic graphs in~$\mathbb{H}^{d+1}$~\cite{krioukov2010hyperbolic}; \emph{USW}: ultrasmall worlds; \emph{SW}: small worlds; \emph{LW}: large worlds; \emph{ZC}: zero clustering at $n\to\infty$; \emph{PC}: positive clustering at $n\to\infty$. If $f(x)$ grows slower or faster than logarithmically, then the networks are in the first and last rows, respectively. The $\gamma=\infty$ case is the homogeneous ensemble in Table~\ref{tab:homo}.
\label{tab:hetero}
}
\end{table}

The calculation of the link length distribution~$l(x)$ in this case in Sec.~\ref{app4} yields $l(x) \sim x^{-\delta}$ with $\delta=\beta-d+1$ if $\beta<d(\gamma-1)$, and $\delta=d(\gamma-2)+1$ otherwise. Following the same logic behind the necessary conditions for small worldness as in the homogeneous case, which says that the networks can be small worlds only if $\xi=1/(\delta-1)>1/d$, we conclude that small worlds are possible if $\beta<2d$ or $\gamma<3$, or both. The networks are necessarily large worlds if $\beta>2d$ and $\gamma>3$. A more detailed analysis proves that these necessary conditions for small worldness are also sufficient~\cite{deprez2015inhomogeneous,norros2006conditionally,esker2008universality}. In fact, the qualitative clustering/small-worldness yes/no diagram for any $\gamma>3$ is exactly the same as in Table~\ref{tab:homo} for the homogeneous ensemble with $f(x)=\ln{x}$ and $l_{\inf}=l_{\sup}=1$. If $\gamma<3$, then our networks are worlds that are not only small but also ultrasmall, regardless of the value of~$\beta$~\cite{bringmann2016average,komjathy2019explosion}.

\begin{figure}[t]
	\centering
	\includegraphics[width=1\linewidth]{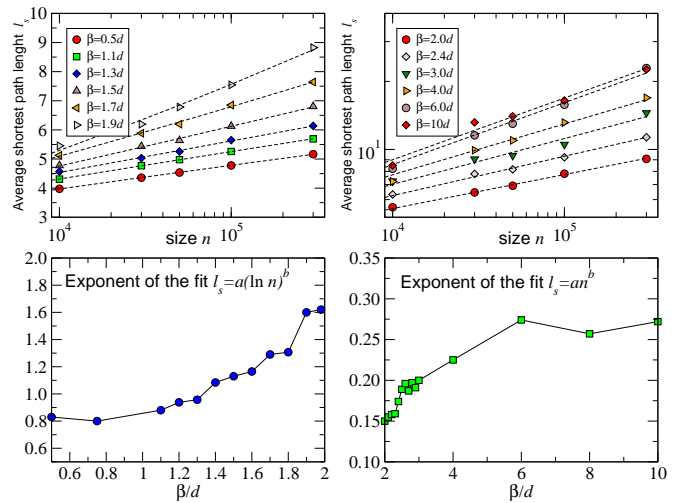}
	\caption{{\bf The average shortest path length~$l_s$ in the heterogeneous spatial networks as a function of the network size~$n$.}
The settings are the same as in Fig.~\ref{fig:1}, except that the networks are heterogeneous~\eqref{eq:fda} with Pareto~$\rho(\kappa)$~\eqref{eq:pareto} and $\gamma=3.5$.
}
	\label{fig:2}
\end{figure}

Table~\ref{tab:hetero} collects all the results, and Fig.~\ref{fig:2} and Figs.~\ref{fig:S2}--\ref{fig:S1} in Sec.~\ref{app5} confirm them in simulations. One sees in Fig.~\ref{fig:2} that if $\gamma>3$, then the simulation results are qualitatively similar to the homogeneous case, except that they are noisier, and the values of exponent~$b$ are significantly smaller. If the network size is small, such small values of~$b$ can be deceiving, making these large worlds appear as small worlds.

We finally remark that the homogeneous ensemble is the $\gamma\to\infty$ limit of the heterogeneous one, because at $\gamma\to\infty$ the Pareto distribution~$\rho(\kappa)$ becomes the degenerate distribution~$\delta(\kappa-\kappa_0)$, so that $\rho(\alpha)\to\delta(\alpha+\beta\mu/2)$ recovering~\eqref{eq:fd} from~\eqref{eq:fda}. In the infinite temperature $\beta\to0$ limit, the connection probability~\eqref{eq:fda} is equal to~$1/(e^{\alpha_i+\alpha_j}+1)$, which is the connection probability in the hypergrandcanonical or hypersoft configuration model that defines the unique ensemble of unbiased random graphs whose entropy is maximized across all graphs with a given degree distribution~\cite{hoorn2018sparse}. In the opposite zero temperature $\beta\to\infty$ limit, the ensemble is equivalent to random hyperbolic graphs with a sharp connectivity threshold~\cite{krioukov2010hyperbolic}. Finally, the $\gamma\to\infty,\beta\to0$ limit is ER.

{\bf In summary,} in spatial networks that are either homogeneous or have a finite degree distribution variance, the decay of the connection probability function with distance~$x$ in a space of dimension~$d$ must be between~$\sim x^{-d}$ and~$\sim x^{-2d}$ to yield sparse small worlds with nonzero clustering. If the degree distribution variance is infinite though, then the spatial networks are ultrasmall worlds with any connection probability, and they have nonzero clustering if this probability decays with~$x$ faster than~$x^{-d}$. Small worldness is linked to link energy and the distribution of link lengths. Networks are small worlds if they contain links of all lengths up to the space diameter. Clustering is dictated by the integrability of the connection probability function. If it is not integrable, then it must decay with the networks size~$n$ to let the network be sparse, but then clustering is zero. This is directly related to the important notion of projectivity~\cite{shalizi2013consistency,orbanz2010conjugate}: if the connection probability depends on~$n$, then the network model is not projective, leading to nonlocal effects that cannot be present in any real-world network~\cite{hoorn2018sparse,krioukov2013duality,spencer2017projective}. We thus see that any realistic model of sparse spatial networks must necessarily have nonzero clustering, which is natural.

As a final comment, we have presented spatial network models as hypergrandcanonical ensembles, probabilistic mixtures of grand canonical ones. In the latter ensembles, the constraints under which the ensemble entropy is maximized are clear: the average energy and the average number of particles in the ensemble, that fix the average link length and average degree, or a sequence of expected degrees, respectively. What remains unclear is under what constraints the considered hypergrandcanonical ensembles are entropy maximizers. Are these constraints similar to the grand canonical ones, or are they completely different, perhaps related to the expected number of triangles in the network~\cite{krioukov2016clustering}? In other words, what are the unbiased maximum entropy spatial network models for sparse heterogeneous small worlds with nonzero clustering?

% =================================================================================================
% Acknowledgments
% =================================================================================================
%
\section*{Acknowledgments}
We acknowledge support from a James S. McDonnell Foundation Scholar Award in Complex Systems; the ICREA Academia prize, funded by the Generalitat de Catalunya; Ministerio de Econom\'{\i}a y Competitividad of Spain project no. FIS2016-76830-C2-2-P (AEI/FEDER, UE); the project {\it Mapping Big Data Systems: embedding large complex networks in low-dimensional hidden metric spaces} -- Ayudas Fundaci\'on BBVA a Equipos de Investigaci\'on Cient\'{\i}fica 2017; Generalitat de Catalunya grant No.~2017SGR1064; the NSF Grant No. IIS-1741355; and ARO Grant Nos.~W911NF-16-1-0391 and~W911NF-17-1-0491.

\appendix

\section{Spatial networks as hypergrandcanonical ensembles} 
\label{app1}
Let $\mathcal{G}\{\mathbb{A};P(\mathbb{A})\}$ be an ensemble of networks with adjacency matrices $\mathbb{A}=\{ a_{ij}\}$ and probability measure~$P(\mathbb{A})$. Let also $F_l(\mathbb{A})$ be an arbitrary set of network functions. The canonical ensemble of random graphs that maximize the Gibbs entropy 
\begin{equation}
  S=-\sum_{\mathbb{A}}P(\mathbb{A}) \ln{P(\mathbb{A})}
\end{equation}
under the constraints that the ensemble averages of~$F_l$,
\begin{equation}
  \langle F_l \rangle = \sum_{\mathbb{A}}F_l(\mathbb{A})P(\mathbb{A}),
\end{equation}
are fixed to some values~$\bar{F}_l$, is given by the Boltzmann/Gibbs distribution
\begin{align}\label{eq:gibbs}
P(\mathbb{A})&=\frac{e^{-\sum_l \alpha_l F_l(\mathbb{A})}}{Z},\text{ where}\\
Z&=\sum_{\mathbb{A}} e^{-\sum_l \alpha_l F_l(\mathbb{A})}
\end{align}
is the partition function, and $\alpha_l$ the Lagrange multipliers coupled to the constraints~$\langle F_L \rangle = \bar{F}_l$. The values of~$\bar{F}_l$ determine the values of~$\alpha_l$~\cite{park2004statistical}.

The Gibbs distribution is known as an exponential family distribution in statistics, so that such canonical ensembles are called exponential random graphs there~\cite{robins2007introduction}. This distribution is known to be the unique unbiased distribution: it is proven that given the constraints, this is the unique distribution that encodes all the information contained in the constraints, and more importantly, it does not encode any other information~\cite{shannon1948mathematical,shore1980axiomatic,tikochinsky1984consistent}.

Note that node pairs enumerate the $n\choose2$ particle states~$\{i,j\}$, $i<j$, that particles---that is, links---can occupy. If the graphs are simple and unweighted, then particles are fermions: there can be either zero or one particle at any particle state. If state~$\{i,j\}$ is occupied, then $a_{ij}=1$, and $a_{ij}=0$ otherwise. Different system states then corresponds to different networks~$\mathbb{A}$, and the number of particles in a system state~$\mathbb{A}$ is
\begin{equation}
  M(\mathbb{A})=\sum_{i<j} a_{ij}.
\end{equation}

Suppose now that nodes in these networks are $n$ fixed points in any Riemannian manifold. The coordinates of these points define the distance matrix~$\mathbb{X}=\{x_{ij}\}$ between them on the manifold. Given any function~$f(x)$, we call $\epsilon_{ij}=f(x_{ij})$ the energy of the particle state~$\{i,j\}$. The energy of the system state~$\mathbb{A}$ is then
\begin{equation}
  E(\mathbb{A}) = \sum_{i<j}\epsilon_{ij}a_{ij}=\sum_{i<j}f(x_{ij})a_{ij}.
\end{equation}

Consider now the canonical ensemble defined by just two functions
\begin{align}
  F_0(\mathbb{A}) &= E(\mathbb{A}),\\
  F_1(\mathbb{A}) &= M(\mathbb{A}),
\end{align}
and two constraints
\begin{align}
  \langle E \rangle &= \bar{E},\\
  \langle M \rangle &= \bar{M},\label{eq:M-constraint}
\end{align}
where $\bar{E},\bar{M}$ are given real numbers. Note that this ensemble is a vanilla grand canonical ensemble in statistical physics that maximizes ensemble entropy under the average energy and number of particles constraints. The latter constraint fixes the average number of links and consequently the average degree, while the former constraint fixes the average link length.

Denoting the Lagrange multipliers by
\begin{align}
  \alpha_0 &=\beta, \\
  \alpha_1 &=-\beta\mu,\label{eq:alpha1}
\end{align}
one can check~\cite{park2004statistical} that the ensemble distribution~\eqref{eq:gibbs} can be written as
\begin{align}
  P(\mathbb{A})&=\frac{\prod_{i<j}e^{\beta(\mu -\epsilon_{ij}) a_{ij}}}{Z}\nonumber\\
  &=\prod_{i<j} p_{ij}^{a_{ij}}(1-p_{ij})^{1-a_{ij}},\text{ where} \\
  p_{ij}&=\frac{1}{e^{\beta(\epsilon_{ij}-\mu)} + 1}\label{eq:fdij}
\end{align}
is the connection probability that takes the standard Fermi-Dirac form, and where the values of the chemical potential~$\mu$ and the inverse temperature~$\beta$ determine the average degree and link length, respectively.

In heterogeneous networks, instead of one average degree constraint~\eqref{eq:M-constraint}, we have $n$ per-node constraints
\begin{equation}
  \langle k_i \rangle = \left \langle \sum_{j} a_{ij} \right \rangle = \kappa_i,
\end{equation}
where $\kappa_i$ is any given sequence of expected degrees. Consequently, instead of one Lagrange multiplier~\eqref{eq:alpha1}, we have $n$ Lagrange multipliers~$\alpha_i$. The values of~$\kappa_i$ determine the values of~$\alpha_i$ via the set of equations derived below for a particular case of interest. One can check that the homogeneous Fermi-Dirac connection probability~\eqref{eq:fdij} changes in the heterogeneous case to
\begin{equation}
  p_{ij}=\frac{1}{e^{\beta\epsilon_{ij}+\alpha_i+\alpha_j} + 1}.
\end{equation}

Suppose now that the coordinates of nodes are no longer fixed on the manifold, but that they are random, e.g.\ that they are a binomial or Poisson point process. Then distances~$x_{ij}$ are random as well, and so are energies~$\epsilon_{ij}$. Similarly, in the heterogeneous case, suppose that $\kappa_i$ are no longer fixed either, but also random, e.g.\ sampled from a fixed distribution~$\rho(\kappa)$. The Lagrange multipliers~$\alpha_i$ are then also random. In both the homogeneous and heterogeneous cases the connection probabilities~$p_{ij}$ are now random. Yet since links are still established with the same, albeit random, Fermi-Dirac connection probabilities~\eqref{eq:fdij}, the resulting ensembles with random coordinates (and random expected degrees) are probabilistic mixtures of the grand canonical ensembles defined above. We call these mixtures \emph{hypergrandcanonical ensembles} since energies~$\epsilon_{ij}$ (and degrees~$\kappa_i$) are no longer parameters but hyperparameters instead. These hypergrandcanonical mixtures are conceptually no different from how a (grand)canonical ensemble is itself a probabilistic mixture of microcanonical ensembles~\cite{touchette2015equivalence}.

\section{No degree correlations $\Leftrightarrow f(x)=c \ln{x}$} 
\label{app2}
Here we show that the energy function $f(x)=c\ln{x}$ is the unique one leading to the absence of correlations of expected degrees in the considered ensembles in the thermodynamic limit, so that all degree correlations (if any) are only structural~\cite{boguna2004cut}.

Formally, the absence of expected degree correlations means that the probability distribution $P(\kappa'|\kappa)$ of expected degrees~$\kappa'$ of nodes to which a random node of expected degree~$\kappa$ is connected, does not depend on~$\kappa$. Instead of expected degrees~$\kappa,\kappa'$ and distribution $P(\kappa'|\kappa)$, it is more convenient to work with the corresponding Lagrange multipliers~$\alpha,\alpha'$ and distribution $P(\alpha'|\alpha)$, the latter also independent of~$\alpha$ if there are no expected degree correlations. Using results from~\cite{boguna2003class},
\begin{equation}
P(\alpha'|\alpha)=\frac{\rho(\alpha') F(\alpha+\alpha')}{\int d\alpha'' \rho(\alpha'') F(\alpha+\alpha'')}
\label{transition}
\end{equation}
where $\rho(\alpha)$ is the distribution of~$\alpha$ defined by~$\rho(\kappa)$ given the relations between $\kappa$s and $\alpha$s as documented in the subsequent section, and function~$F$ is defined to be
\begin{equation}
F(\alpha+\alpha')=\int_0^{\infty} \frac{x^{d-1}dx}{1+e^{\beta f(x)+\alpha+\alpha'}}.
\end{equation}
To find under which conditions $P(\alpha'|\alpha)$ is independent of $\alpha$, we differentiate~\eqref{transition} with respect to~$\alpha$ and equate the result to zero to obtain
\begin{equation}
\frac{F'(\alpha+\alpha')}{F(\alpha+\alpha')}=\frac{\int \rho(\alpha'')F'(\alpha+\alpha'')d \alpha''}{\int \rho (\alpha'')F(\alpha+\alpha'')d \alpha''}.
\end{equation}
Since the right hand side of this equation does not depend on $\alpha'$, function~$F$ is of the form $F(x)=a e^{b x}$, with $a$ and $b$ some constants. Define $q(x)\equiv e^{f(x)}$ and $z \equiv e^{-(\alpha+\alpha')/\beta}$ to rewrite the uncorrelatedness condition as
\begin{equation}
\int_0^{\infty} \frac{x^{d-1}dx}{1+(q(x)/z)^{\beta}}=a z^{b \beta}.
\label{identity}
\end{equation}
That is, the network is uncorrelated at the level of hidden variables~$\alpha,\alpha'$ whenever Eq.~\eqref{identity} holds for any value of $z \in \mathbb{R}^+$, with $a$ and $b$ some constants.

\subsection{If $ f(x)=c \ln{x}$, then Eq.~\eqref{identity} holds}

We first notice that the energy function $f(x)=c \ln{x}$ is a sufficient condition for uncorrelatedness, since then Eq.~\eqref{identity} trivially holds with $b \beta=d/c$ and
\begin{equation}
a=\int_0^{\infty} \frac{t^{d-1}dt}{1+t^{c \beta}}.
\end{equation}

\subsection{If Eq.~\eqref{identity} holds, then $f(x)=c \ln{x}$ }

We are next to prove that in the small world regime where $f(x)/\ln{x}\in(l_{inf},l_{sup})$ for $x > X$, $l_{inf}=\liminf_{x\to\infty}f(x)/\ln x$, $l_{sup}=\limsup_{x\to\infty}f(x)/\ln x$, and some constant $X>0$, the assumption that Eq.~\eqref{identity} holds implies that $f(x)=c \ln{x} \; \forall x \in \mathbb{R}^+$. We consider two cases.

\subsubsection{Case $l_{inf}=l_{sup}=c$}

In this case, function $f(x)/\ln x$ has a limit, 
\begin{equation}
\lim_{x \rightarrow \infty} \frac{f(x)}{\ln{x}}=c,
\end{equation}
and since $q(x)= e^{f(x)}$, we have
\begin{equation}
\lim_{x \rightarrow \infty} \frac{q(x)}{x^c}=1.
\end{equation}
If we define $\hat{q}(x)\equiv q(x)/x^c$, Eq.~\eqref{identity} can be written as
\begin{equation}
z^{\frac{d}{c}}\int_0^{\infty} \frac{t^{d-1}dt}{1+t^{c \beta} [\hat{q}(z^{\frac{1}{c}}t)]^{\beta}}=a z^{b \beta}.
\label{identity2}
\end{equation}
If this equation holds for all values of $z\in \mathbb{R}^+$, then the integral in it must be a power of $z$ for any $z$ including $z\gg1$. Let us split this integral in two:
\begin{equation}
\int_0^{\frac{x_c(\epsilon)}{z^{1/c}}} \frac{t^{d-1}dt}{1+t^{c \beta} [\hat{q}(z^{\frac{1}{c}}t)]^{\beta}}+\int_{\frac{x_c(\epsilon)}{z^{1/c}}}^{\infty} \frac{t^{d-1}dt}{1+t^{c \beta} [\hat{q}(z^{\frac{1}{c}}t)]^{\beta}},
\label{eq:split}
\end{equation}
where $x_c(\epsilon)$ is such that for any $x>x_c(\epsilon)$ we have that $|\hat{q}(x)-1|<\epsilon$. We thus see that function $\hat{q}$ is bounded in the integration domain of the second integral in Eq.~\eqref{eq:split}. This implies that
\begin{equation}
\lim_{z \rightarrow \infty} \int_{\frac{x_c(\epsilon)}{z^{1/c}}}^{\infty} \frac{t^{d-1}dt}{1+t^{c \beta} [\hat{q}(z^{\frac{1}{c}}t)]^{\beta}}=\int_0^{\infty} \frac{t^{d-1}dt}{1+t^{c \beta} },
\end{equation}
which is a constant independent of $z$. At the same time, the limit $z\rightarrow \infty$ of the first integral in Eq.~\eqref{eq:split} is zero because the domain of integration goes to zero and the integrand does not diverge at zero. Combining all these observations with Eq.~\eqref{identity2} we conclude that $b \beta=d/c$ and that
\begin{equation}
a=\int_0^{\infty} \frac{t^{d-1}dt}{1+t^{c \beta} [\hat{q}(z^{\frac{1}{c}}t)]^{\beta}}=\int_0^{\infty} \frac{t^{d-1}dt}{1+t^{c \beta} } \; \forall z\in \mathbb{R}^+.
\end{equation}
This is possible only if $\hat{q}(x)=1$, and hence $f(x)=c\ln{x}$. 

\subsubsection{Case $l_{inf} \neq l_{sup}$}

Let us assume now that $l_{inf}$ and $l_{sup}$ are both positive and finite but not necessarily equal. The condition for uncorrelatedness in Eq.~\eqref{identity} implies that there must exist a value of $b$ such that the limit
\begin{equation}
\lim_{z \to \infty} z^{-b \beta} \int_0^{\infty} \frac{x^{d-1}dx}{1+(q(x)/z)^{\beta}}
\end{equation}
exists. However, if $f(x)/\ln{x}$ is squeezed between $l_{inf}$ and $l_{sup}$ at $x\gg1$, then this integral is squeezed between $z^{\frac{d}{l_{sup}}}$ and $z^{\frac{d}{l_{inf}}}$ at $z \gg 1$, and the limit does not exist, so that we arrive at a contradiction. We thus conclude that the only possibility is that $l_{inf}=l_{sup}=c$, so that $f(x)=c \ln{x}$.

Finally we remark that $c$ can be always set to $1$ by a proper choice of energy units.

\section{Relations between $\kappa$ and $\alpha$} 
\label{app3}
Here, we derive these relations for the heterogeneous hypergrandcanonical ensemble with the energy function $f(x)=\ln{x}$, the Poisson point process of intensity~$1$ in~$\mathbb{R}^d$, and any distribution of expected degrees~$\rho(\kappa)$. The cases with $\beta>d$ and $\beta<d$ must be considered separately.

\subsection{Case $\beta>d$}  

In this case, thanks to the integrability of the connection probability w.r.t.\ the spatial distance, we can work directly in the thermodynamic limit in $\mathbb{R}^d$. Since the space is homogeneous we assume without loss of generality that a node with variable $\alpha$ is at the origin, and we want to calculate its expected degree $\kappa$. It is convenient to work in spherical coordinates in $\mathbb{R}^d$, in which the volume element is
\begin{equation}
dV_{\mathbb{R}^d}=r^{d-1}dr\,dV_{\mathbb{S}^{d-1}},
\end{equation}
where $dV_{\mathbb{S}^{d-1}}$ is the volume element on the unit $(d-1)$-sphere whose volume is
\begin{equation}
  S_{d-1}=\int dV_{\mathbb{S}^{d-1}} = \frac{2 \pi^{\frac{d}{2}}}{\Gamma\left( \frac{d}{2}\right)}.
\end{equation}
The expected degree $\kappa$ of our node is then
\begin{equation}
\kappa=S_{d-1}\int \rho(\alpha')d\alpha' \int_0^{\infty} \frac{r^{d-1}dr}{1+\left(r e^{\frac{\alpha+\alpha'}{\beta}}\right)^{\beta}}.
\end{equation}
Changing variables we simplify this to
\begin{align}
\kappa&=S_{d-1} \langle e^{-\frac{\alpha d}{\beta}}\rangle e^{-\frac{\alpha d}{\beta}} \int_0^{\infty} \frac{t^{d-1}dt}{1+t^{\beta}}\nonumber\\
&=S_{d-1} \langle e^{-\frac{\alpha d}{\beta}}\rangle e^{-\frac{\alpha d}{\beta}} \frac{\pi}{\beta \sin{\frac{d \pi}{\beta}}}.\label{kappa:beta>d}
\end{align}
By taking the average of Eq.~\eqref{kappa:beta>d}, we find the relation between the term $\langle e^{-\frac{\alpha d}{\beta}}\rangle$ and the average degree $\langle k \rangle$, which plugged again in Eq.~\eqref{kappa:beta>d} leads to the final relation between $\alpha$ and $\kappa$:
\begin{align}
\alpha&=-\frac{\beta}{d} \left(\ln{\kappa}+\frac{1}{2} \ln{\hat{\mu}} \right) \mbox{, where }\\
\hat{\mu}&= \frac{\beta\sin{\frac{d \pi}{\beta}}}{\pi S_{d-1} \langle k \rangle} = \frac{\beta \Gamma(\frac{d}{2})\sin{\frac{d \pi}{\beta}}}{2 \pi^{1+\frac{d}{2}} \langle k \rangle}.
\end{align}

This result implies that the edge-state energy $\epsilon_{ij}$ and chemical potential $\mu$ in the ensemble are given by
\begin{align}
\epsilon_{ij}&=\ln{\left[\frac{x_{ij}}{\left(\kappa_i \kappa_j \right)^{\frac{1}{d}}}\right]},\\
\mu&=\frac{1}{d}\ln{\hat{\mu}}.
\end{align}
The connection probability can then be written as
\begin{equation}\label{eq:plt}
p(x_{ij},\kappa_i,\kappa_j)=\frac{1}{1+\left[ \frac{x_{ij}}{(\hat{\mu} \kappa_i \kappa_j)^{\frac{1}{d}}} \right]^{\beta}}\sim\left(\hat{\mu}\kappa_i\kappa_j\right)^{\beta/d}x_{ij}^{-\beta}.
\end{equation}

\subsection{Case $\beta<d$}  

In this case, the connection probability is not integrable w.r.t.\ distance, so that we have to take the finite size effects into account. This implies that the answer depends on a particular choice of the manifold family. Yet we remind that our general settings are such that for any $n$ the manifold volume is $n$, so that
\begin{equation}
n= V_d R^d,
\end{equation}
where $R$ is the linear size of the manifold, while $V_d$ is its volume at $R=1$. For example, if the manifold is a $d$-torus, then $R$ is its side length and $V_d=1$. If it is a $d$-sphere, then $R$ is its radius, and $V_d$ is the volume $S_d$ of the unit $d$-sphere:
\begin{equation}
V_d=S_d=\frac{2 \pi^{\frac{d+1}{2}}}{\Gamma\left( \frac{d+1}{2}\right)}.
\end{equation} 
We consider the case with the $d$-sphere for concreteness.

Since the $d$-sphere is homogeneous, we assume without loss of generality that a node with variable~$\alpha$ is at its north pole. The volume element on the $d$-sphere with $\theta$ the polar angle is
\begin{equation}
dV_{\mathbb{S}^d}=\sin^{d-1}d\theta\,dV_{\mathbb{S}^{d-1}},
\end{equation}
so that the expected degree of our node is
\begin{equation}
\kappa=S_{d-1}R^d \int \rho(\alpha')d\alpha' \int_0^{\pi} \frac{\sin^{d-1}\theta d\theta}{1+\left(R \theta e^{\frac{\alpha+\alpha'}{\beta}}\right)^{\beta}}.
\label{kappa:beta<d}
\end{equation}
For $R \gg 1$ and $\beta<d$ the integral in the last expression can be approximated as
\begin{equation}
\int_0^{\pi} \frac{\sin^{d-1}\theta d\theta}{1+\left(R \theta e^{\frac{\alpha+\alpha'}{\beta}}\right)^{\beta}}\sim \frac{\pi^{d-\beta}}{d-\beta} R^{-\beta} e^{-(\alpha+\alpha')}.
\end{equation}
Using this expression in Eq.~\eqref{kappa:beta<d}, we conclude that for $n\gg 1$ the relation between $\alpha$ and the expected degree $\kappa$ is given by 
\begin{align}
\alpha&=-\left(\ln{\kappa}+\frac{1}{2}\ln{\hat{\mu}}\right)\text{, where}\\ 
\hat{\mu}&=
\frac{d-\beta}{\pi^{d-\beta}}\frac{1}{S_{d-1} \langle k \rangle}\left(\frac{S_d}{n}\right)^{1-\frac{\beta}{d}}\nonumber\\
&=\frac{(d-\beta)\Gamma(\frac{d}{2})}{2 \pi^{\frac{3d}{2}-\beta} \langle k \rangle n^{1-\frac{\beta}{d}}} \left[ \frac{2 \pi^{\frac{d+1}{2}}}{\Gamma(\frac{d+1}{2})} \right]^{1-\frac{\beta}{d}}.
\end{align}

The connection probability is then
\begin{equation}\label{hrgg}
p(x_{ij},\kappa_i,\kappa_j)=\frac{1}{1+\frac{x_{ij}^{\beta}}{\hat{\mu}\kappa_i \kappa_j}}\sim \hat{\mu}\kappa_i\kappa_jx_{ij}^{-\beta}.
\end{equation}
This connection probability depends on~$n$ and tends to zero as $\sim 1/n^{1-\beta/d}$ since so does~$\hat{\mu}$.
We also note that it cannot be written as a Fermi-Dirac distribution function, meaning that the energy of edges cannot be defined in the case $\beta<d$.

\section{The link length distribution~$l(x)$} 
\label{app4}
Here, we calculate the tail of this distribution in the heterogeneous hypergrandcanonical ensemble with the energy function $f(x)=\ln{x}$, the Poisson point process of intensity~$1$ in~$\mathbb{R}^d$, and Pareto~$\rho(\kappa)\sim\kappa^{-\gamma}$ with~$\gamma>2$.

If $\beta<d$, then according to~\eqref{hrgg}, the distribution of link lengths is given by
\begin{align}
  l(x) &\sim\iint p(x,\kappa,\kappa')\,x^{d-1} \rho(\kappa)d\kappa\,\rho(\kappa')d\kappa' \nonumber\\
  &\sim x^{-\beta+d-1} \left[\int \kappa\,\rho(\kappa)d\kappa\right]^2 \sim x^{-(\beta-d+1)},
\end{align}
since the Pareto distribution has a finite mean if~$\gamma>2$.

If $\beta>d$, then according to~\eqref{eq:plt}, the expression for~$l(x)$ becomes
\begin{align}
  l(x) &\sim x^{-\beta+d-1} \left[\int\kappa^{\beta/d}\rho(\kappa)d\kappa\right]^2 \nonumber\\
  &\sim x^{-\beta+d-1}\left[\int\kappa^{-\gamma+\beta/d}d\kappa\right]^2,
\end{align}
where the $\kappa$-integral is finite if $\beta<d(\gamma-1)$, so that $l(x)$ is still~$\sim x^{-(\beta-d+1)}$.

If $\beta>d(\gamma-1)$, then the $\kappa$-integral in the last equation is infinite, so that slightly more care is needed to derive the scaling of~$l(x)$ with~$x$. Specifically, for large~$\beta$ the exact expression for the connection probability~$p(x,\kappa,\kappa')$ in~\eqref{eq:plt} can be approximated by~1 for $x/(\hat{\mu}\kappa\kappa')^{1/d}<1$, and by~$0$ otherwise, in which case we get
\begin{align}
  l(x) &\sim x^{d-1} \iint^{x^d} \rho(\kappa)d\kappa\,\rho(\kappa')d\kappa' \sim x^{-[d(\gamma-2)+1]}.
\end{align}

Collecting all the cases,
\begin{align}
  l(x) &\sim x^{-\delta}, \text{ where }\nonumber\\
  \delta &=
  \begin{cases}
    \beta-d+1, & \mbox{if } \beta<d(\gamma-1) \\
    d(\gamma-2)+1, & \mbox{if } \beta>d(\gamma-1).
  \end{cases}
\end{align}
Note that the link length distribution~$l(x)$ is well-defined in the $n\to\infty$ limit only if $\delta>1$, i.e.\ if $\beta>d$ corresponding to the nonzero clustering regime.

\section{Simulation results} 
\label{app5}
\begin{figure}
	\centering
	\includegraphics[width=0.85\linewidth]{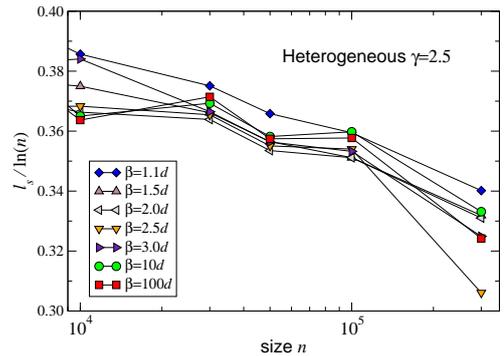}
	\vspace{-0.5cm}\caption{{\bf The average shortest path length~$l_s$ in the heterogeneous spatial networks with $\gamma=2.5$ as a function of the network size~$n$.}
This function is measured for different values of the inverse temperature $\beta$ in the connection probability~(\ref{eq:plt},\ref{hrgg}) with Pareto~$\rho(\kappa)$ with $\gamma=2.5$ used to generate random networks on the $d=2$-dimensional sphere of area~$n$. The average degree in all these networks is fixed to $\langle k \rangle=10$ by the appropriate choice of the chemical potential~$\hat{\mu}$, and the results are averaged over 10 random network realizations for each data point. The functions $l_s(n)$ are divided by $\ln{n}$ to highlight the ultrasmall world distance scaling.
}
	\label{fig:S2}
\end{figure}

Figure~\ref{fig:S2} shows the average shortest path length in the heterogeneous case with $\gamma=2.5$ as a function of $n$. Due to the relatively small network sizes in the simulations, it is not possible to detect the $\ln{\ln{n}}$ ultrasmall world distance scaling that holds for any~$\gamma<3$ and any~$\beta$. However one can see that the average shortest path length grows slower than $\ln{n}$ for any $\beta$.

\begin{figure}
	\centering
	\includegraphics[width=1\linewidth]{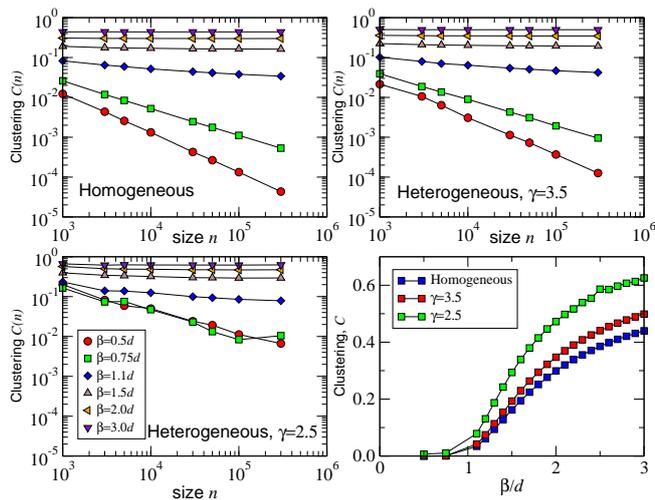}
	\vspace{-0.2cm}\caption{{\bf The average local clustering coefficient in homogeneous and heterogeneous spatial networks as a function of the network size~$n$ for different values of inverse temperature~$\beta$ and power-law exponent~$\gamma$.} The space is the $d=2$-sphere of area~$n$. The average degree is fixed to $\langle k \rangle=10$. The results are averaged over 10 random network realizations for each data point. The bottom right panel shows the average clustering as a function of $\beta/d$ for the largest network size $n=3 \times 10^5$.
}
	\label{fig:S1}
\end{figure}

Figure~\ref{fig:S1} shows the simulation results for the local clustering coefficient, averaged over nodes of degrees greater than 1, as a function of network size $n$ for different heterogeneities and values of $\beta$. The simulations confirm the continuous transition from the limiting zero to nonzero clustering at $\beta = d$.

%\bibliographystyle{apsrev4-1}
%\bibliography{geometry3}

%merlin.mbs apsrev4-1.bst 2010-07-25 4.21a (PWD, AO, DPC) hacked
%Control: key (0)
%Control: author (72) initials jnrlst
%Control: editor formatted (1) identically to author
%Control: production of article title (-1) disabled
%Control: page (0) single
%Control: year (1) truncated
%Control: production of eprint (0) enabled
%

\end{document}